\documentclass[12pt]{article} 
\usepackage{epsfig} 
 1 
\def\as{\alpha_{\rm s}}
\def\aem{\alpha_{\rm em}}

\def\dd{\partial}

\def\percent{{\%\ }}

\newcommand{\half}{\mbox{\small $\frac{1}{2}$}}

\def\lsim{\;\raisebox{-.4ex}{\rlap{$\sim$}} \raisebox{.4ex}{$<$}\;}

\def\gev{{\,\mbox{GeV}  }}
\def\gevs{{\,\mbox{GeV}^2}}

\def\l{\lambda}

\def\({\left(}
\def\){\right)}

\def\citenum#1{{\def\@cite##1##2{##1}\cite{#1}}}
\def\citea#1{\@cite{#1}{}}

\def\l1vt{\vec{l_{1\perp}}}

\def\rt{r_{\perp}}
\def\rhot{\rho_{\perp}}
\def\bt{b_{\perp}}

\def\abst{|t|}

\def\jol1{$J_0(\,l_{1\perp}\,r_{\perp}\,)$}

\def\ie{\hbox{\it i.e.}}        

\def\eg{\hbox{\it e.g.}}        

\def\etal{\hbox{\it et al.}}

\def\half{{1\over 2}}

\def\beq{\begin{equation}}
\def\eeq{\end{equation}}
\def\bea{\begin{eqnarray}}
\def\eea{\end{eqnarray}}
\def\eq#1{{\mbox{Eq.\hspace{1mm}(\ref{#1})}}}

\def\fig#1{{\mbox{Fig.\hspace{1mm}\ref{#1}}}}

\def\scrbox#1{\mbox{\scriptsize #1}}

\def\hfill{\hspace*{\fill}}

\def\npb#1#2#3{    {\it Nucl.\ Phys.\ }{\bf B#1} (#2) #3}

\def\plb#1#2#3{    {\it Phys.\ Lett.\ }{\bf B#1} (#2) #3}
\def\prd#1#2#3{    {\it Phys.\ Rev.\ }{\bf D#1} (#2) #3}
\def\prep#1#2#3{   {\it Phys.\ Rep.\ }{\bf #1} (#2) #3}

\def\rmp#1#2#3{    {\it Rev.\ Mod.\ Phys.\ }{\bf #1} (#2) #3}
\def\zpc#1#2#3{    {\it Z.\ Phys.\ }{\bf C#1} (#2) #3}

\def\yf#1#2#3{     {\it Yad.\ Fiz.\ }{\bf #1} (#2) #3}
\def\sjnp#1#2#3{   {\it Sov.\ J.\ Nucl.\ Phys.\ }{\bf #1} (#2) #3}
\def\jetp#1#2#3{   {\it Sov.\ Phys.\ }{JETP }{\bf #1} (#2) #3}

\def\epj#1#2#3{    {\it Eur.\ Phys.\ J.\ }       {\bf #1} (#2) #3}

\newcommand{\email}[1]{${\!}^{\scrbox{#1)}}$}

\newcommand{\Qbar}{\overline{Q}}
\newcommand{\Qbars}{\overline{Q}^{\:2}}

\newcommand{\sighat}{\hat\sigma}

\def\df2dlnq2{\dd{F_2}/\dd\log{Q^2}}
\def\dlnxg{\dd\log{xG}/\dd\log(1/x)}
\newcommand{\bslopeI}{\left(1-e^{-\half\Omega_q}\right)}
\newcommand{\bslopeII}{
\frac{C_F}{\pi^2}\as\rt^2\int\frac{dx}{x}
\int_{\rhot>\rt}\frac{d^2\rhot}{\rhot^4}\left(1-e^{-\half\Omega_g}\right)
}

\relax
\begin{document}
\begin{titlepage}
\noindent
\begin{flushright}
\parbox[t]{14em}{
TAUP 2691/2001\\
{\tt hep-ph/0110256}\\
To appear in {\it Phys.\ Lett.\ }{\bf B}
}
\end{flushright}

\vspace{1cm}
\begin{center}
  {\Large \bf Momentum Transfer Dependence of the 
Differential Cross Section for
    ${\mathbf J/\psi}$ Production}
  \\[4ex]

\begin{minipage}{0.6\textwidth}
{\large E. ~G O T S M A N \email{1}, \hspace{\fill}
        E. ~L E V I N     \email{2}, \vspace{2ex}\\
        U. ~M A O R       \email{3}  \hfill and \hfill 
        E. ~N A F T A L I \email{4}}
\end{minipage}
\footnotetext{\email{1} Email: gotsman@post.tau.ac.il .}
\footnotetext{\email{2} Email: leving@post.tau.ac.il .}
\footnotetext{\email{3} Email: maor@post.tau.ac.il .}
\footnotetext{\email{4} Email: erann@post.tau.ac.il .}
\\[4.5ex]
{\it  School of Physics and Astronomy}\\
{\it  Raymond and Beverly Sackler Faculty of Exact Science}\\
{\it  Tel Aviv University, Tel Aviv, 69978, ISRAEL}\\[4.5ex]

\end{center}
~\,\,
\vspace{1cm}

{\samepage {\large \bf Abstract:}} We discuss the $\abst$ dependence of
 $J/\psi$ production in the region of $0<\abst\lsim M^2_{\psi}$. The forward
 slope of the elastic differential cross section is calculated assuming a
 dipole-type dependence for the gluon ladder-proton form factor. The $\abst$
 dependence of $d\sigma/dt$ is obtained using DGLAP evolution both for the
 elastic channel at low $\abst$ and for the inelastic channel at large values
 of $\abst$, up to $\abst\simeq M^2_{\psi}$. Results are presented and
 compared with the relevant experimental data.

\end{titlepage}
\section{Introduction}

Measurements of high energy exclusive production of $J/\psi$ meson in
 photon-proton collisions serve as an important testing ground to quantify
 the ``hard'' physics of pQCD in the limit of a small scaling variable $x$.
 A process is considered to be ``hard'' if a large momentum scale is
 involved, so that the leading contribution of the hadronic fluctuation of
 the photon is a small perturbative $q {\bar q}$ pair.  This large momentum
 scale may be a heavy quark mass, large virtuality of the photon $Q^2$ or
 large four momentum transfer squared at the proton vertex ($-t$).  For the
 diffractive photoproduction of $J/\psi$ this hard scale is provided by the
 charm quark mass.  In our recent letter \cite{GLMFNpsi} we presented a
 screened model which yielded impressive agreement with the rich experimental
 data on high energy $J/\psi$ photoproduction.  The model includes
 contributions of the off diagonal (skewed) gluon distributions
 \cite{OFFDIAGONAL}, as well as the real part of the production amplitude. In
 addition, we addressed the issue of Fermi motion of the heavy quark within
 the quarkonium system \cite{FERMI}, which provides further suppression of
 the final cross section.  We argued that on one hand this effect is
 sensitive to the charm quark mass, while on the other hand, in the relevant
 kinematical region of the experimental data, it has a rather mild energy
 dependence. An obvious conclusion of the above argument is that with the
 current theoretical uncertainties regarding the Fermi motion suppression,
 the benefits resulting from a phenomenological model would be rather small.
 The correction due to Fermi motion has been estimated as an overall
 suppression factor of approximately 30\%.

In the present work, we present an analysis for the $t$ dependence of the
 $J/\psi$ differential cross section which is compatible with our recent
 investigation of the integrated cross section.  To compare with the
 experimental data we need to know $B$ - the $J/\psi$ forward slope of the
 differential cross section.  In \cite{GLMFNpsi} we assumed a weak dependence
 of $B$ on $x$, which was calculated by averaging the square of the impact
 parameter $b$, using our screening correction (SC) formalism
 \cite{GLMV2}. This input assumption is not sufficient for a detailed $t$
 dependence study of $J/\psi$ differential cross sections as the presently
 available preliminary experimental data on $B$ \cite{Bslope} show a steeper
 energy dependence than our prediction.  The goals of the present improved
 version of our model are, thus, to reproduce the value as well as the energy
 dependence of $B$ while maintaining the quality of the results achieved in
 \cite{GLMFNpsi}.

In our previous study, a $b$-space Gaussian was assumed for $S(b)$, the gluon
 ladder-proton non perturbative form factor \cite{GLMFNpsi,GLMV2,GLMN1}.
 After a careful check, it appears that the integrated cross sections
 \cite{GLMFNpsi} and nucleon structure function \cite{GLMN1,GLMNslope2} are
 not sensitive to the choice of $S(b)$.  The impact parameter dependence
 plays an important role, when calculating screening corrections to
 heavy nuclei structure functions, where a Wood-Saxon like \cite{Enge}
 profile function has been used \cite{THERA}, or for the $J/\psi$
 differential photoproduction cross sections \cite{MSM} which is the main
 goal of the present study.  As we shall demonstrate, the experimental data
 \cite{Bslope} for the forward differential cross section slope can be well
 reproduced with the observation that the energy dependence of $B$ is
 sensitive to the choice of $S(b)$.  It appears that the Fourier transform of
 the dipole electromagnetic form factor, is the preferred profile for
 estimating the screening corrections to $B$.

The momentum transfer $t$ at the proton vertex controls the size of the
 interacting system, which emits the gluon ladder.  The $t$ dependence of
 hard processes can, thus, be separated into three kinematical zones.  The
 low $\abst$ region, where the transverse size of the interacting system is
 of the order of $1/Q_0$, where $Q^2_0$ is the input scale for the
 DGLAP\cite{DGLAP} evolution ($Q^2_0\sim 1\gevs$). This region, in which
 $0<\abst<Q_0^2$, has been explored experimentally, mainly via elastic
 processes \cite{H1lowt}.  The hard scale in this region is determined by the
 c-quark mass $m^2_c\approx M_{J/\psi}^2/4 $.  The second region is the
 intermediate region $Q_0^2<\abst<M_{\psi}^2$. We suggest that the physics of
 this region is similar to the physics of the small $\abst$ region, \ie the
 process is still dominated by large logarithms due to momentum integration
 over loops in the ordered gluon ladder.  However, the scale of the initial
 conditions of the DGLAP evolution is determined by $t$, as it becomes of
 comparable size to $Q^2_0$.  More specifically, the process is dominated by
 $\log{(M_{J/\psi}^2/t)}$. In the intermediate region, the vector meson is
 produced mostly quasi elastically, while the proton dissociates into a
 diffractive mass.  Note that the diffractive $t$ slope is considerably
 smaller than the elastic one.  In the present study we compare our
 calculations in this region with preliminary experimental data which were
 read off figures given in Refs.~\cite{H1larget,ZEUSlarget}. As $\abst$
 increases, the hard process behavior is dominated by the value of $\abst$,
 the $\abst$ dependence is inverted \cite{FORSHAW} and the momenta on the
 gluon ladder are no longer strongly ordered.  Consequently, predictions for
 the cross section cannot be described using conventional DGLAP evolution.
 This third region of large $\abst$ has not been explored experimentally.

These three regions of $\abst$ were first discussed in \cite{FORSHAW} by
 Forshaw and Ryskin and by Bartels, Forshaw, Lotter and Wuesthoff in the
 framework of the BFKL equation \cite{BFKL}, where it was stated that each of
 the above regions corresponds to different physics.  In a later publication
 \cite{FORSHAW2}, a BFKL calculation of vector meson production, and in
 particular of $J/\psi$ production, was presented and a good description of
 experimental data was obtained.

In this paper, we suggest a description of the second kinematic region of
 intermediate $\abst$ ($Q_0^2<\abst\lsim M_{\psi}^2$) using the DGLAP
 \cite{DGLAP} approach, this yields simple and transparent formulae which
 have explicit matching to the region of small $\abst$.

Our paper is organized as follows: in section \ref{sec:B} we calculate the
 forward slope for different profiles and compare with the experimental
 data. In sections \ref{sec:low-t} and \ref{sec:high-t} we calculate the
 differential cross sections for exclusive $J/\psi$ production both in the
 elastic and the inelastic channels, respectively. Our conclusions are given
 in section 5.

\section{\label{sec:B}Forward Slope for Differential Cross Section}

In this section we calculate the $B$-slope for $J/\psi$ production, including
 screening corrections for up to one extra gluon emission. Our formalism for
 calculating the screening corrections \cite{SC} is based on the iteration of
 the non-linear evolution equation \cite{GLR} for the imaginary part of the
 elastic amplitude for a dipole to scatter off a hadron target, which is
 valid in the whole kinematic region, including the low $x$ region
 \cite{Kovchegov}.

The interaction of a dipole with the target is realized through an exchange
 of gluons. In the target rest frame, one can use Gribov's factorization in
 the context of a dipole-nucleon interaction: a dipole emits a soft gluon
 which develops into a ladder by successive emissions of small $x$
 gluons. Subsequently, the gluon ladder interacts with the target. At high
 energies, the transverse size of an interacting dipole does not change
 during the QCD interaction, and therefore dipoles are good degrees of
 freedom.  In the limit of large number of colors, a leading logarithmic soft
 gluon wave function is equivalent to a dipole wave function, and thus, the
 interaction between a dipole and a target can be treated as a process which
 is subsequent to a transition of a dipole into two dipoles.  Consider a
 decay of a dipole with transverse size $\rt$ into two dipoles as shown in
 \fig{fig:dipole}.  The probability for this decay is given by the square of
 the wave function of the dipole.  At large $N_c$ one can view this decay as
 an emission of a zero transverse size gluon. In this framework, the cross
 section for the virtual photon-nucleon interaction is written as
\begin{equation}
\sigma_{tot}(\gamma^*p) = \int d^2\rt \int dz \left|
  \Psi^{\gamma^*}(Q^2;\rt,z)\right|^2 \sigma_{dipole}(\rt,z)
\end{equation}
 where the wave functions of the virtual photon are well known, in particular
 the transverse wave function is given by the Bessel function $K_1$.  In our
 model, the dipole-nucleon cross section consists of two contributions:

\begin{enumerate}
 \item The first, denoted $\sighat^q(\rt)$, is the percolation of a
$q\bar{q}$ pair through the target, without taking into account the extra
gluon, and it is calculated using a multi gluon ladder exchange.  This
contribution can be expressed as a product of the dipole wave function and
the cross section for the dipole to scatter off the nucleon. In impact
parameter space, $\sighat^q$ can be written in the form \cite{SC,MU90}
\begin{equation}\label{eq1}
\sighat^q(\rt) \propto \int d^2\bt d^2\rt K_1(\Qbar\rt)\bslopeI,
\end{equation}
where $\Qbar=(Q^2+ M_{\psi}^2)/4$, $x=(Q^2+ M_{\psi}^2)/W^2$ and the
opacity $\Omega_q=\Omega_q(b,\rt,x)$ is defined as
\begin{equation}\label{omegaq}
\Omega_q=\Omega_q(b,\rt,x)= 
\frac{\pi^2}{3}\rt^2\as\(\frac{4}{\rt^2}\)xG\(x,\frac{4}{\rt^2}\)S(\bt).
\end{equation}
 The $\bt$ dependence of $\Omega_q$ is expressed through the profile function
 $S(\bt)$, which we shall discuss later. Note that the argument of the Bessel
 function should be $a\rt$, where $a^2=z(1-z)Q^2+M_{\Psi}^2/4$, however, for
 heavy quarks, we use the approximation $z=\half$.

\begin{figure}
\begin{center}
\epsfig{file=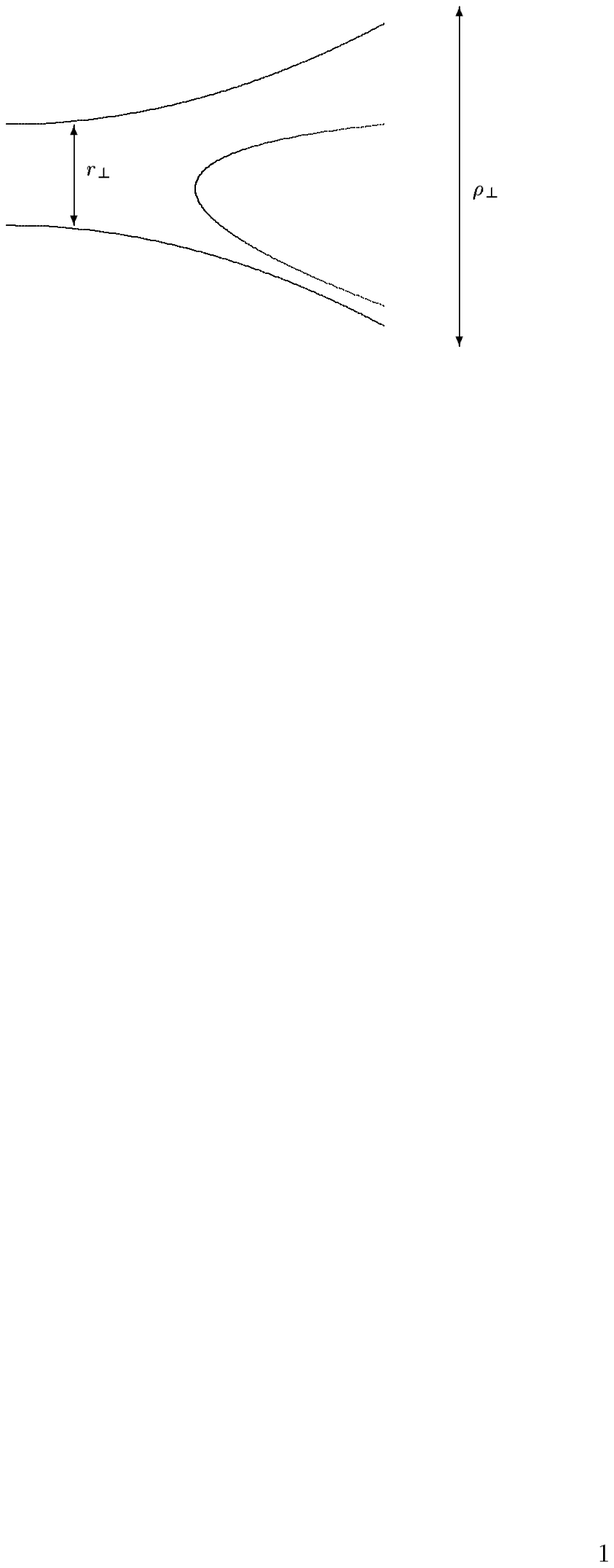,width=8cm,bbllx=66,bblly=613,bburx=267,bbury=756}
\end{center}
  \caption[]{\parbox[t]{0.80\textwidth}{\small
A decay of one dipole with transverse 
size $\rt$ into two dipoles. In the case where two of the produced quarks
are with small transverse separation, they can be view as a zero transverse
size gluon, hence the final sate is a $q\bar{q}g$ state.}}
\label{fig:dipole}
\end{figure}
 \item For the second contribution, denoted $\sighat^g(\rt)$ below, we
consider an additional gluon in the hadronic state which percolates through
the target.  We assume that the parent dipole whose transverse size is $\rt$,
is larger than the produced dipole whose transverse size is $\rhot$ (see
\fig{fig:dipole}). The second contribution can be written in the form
\cite{SC,MU90}:
\begin{equation}\label{eq2}
\sighat^g(\rt) \propto\int d^2\bt d^2\rt K_1(\Qbar\rt) \bslopeII ,
\end{equation}
with $\Omega_g(b,\rt,x)=\frac{9}{4}\Omega_q(b,\rt,x)$.
\end{enumerate}

The forward slope $B$ is given by \cite{Block}:
 \begin{equation}\label{eq:av}
B= \half\langle \bt^2 \rangle = \frac{\int \bt^2 d\bt^2 S(\bt)}{2\int d\bt^2
S(\bt)},
\end{equation}
 where the factor of one half is introduced since $B$ is measured for cross
 sections rather than amplitudes.  To calculate the forward slope $B$, we need
 to average over the impact parent $\bt$, using the sum of (\ref{eq1}) and
 (\ref{eq2}) as weights,
\begin{equation}\label{bslope}
B=\frac{\int d^2\bt d^2\rt \bt^2 K_1(\Qbar\rt) \left[\bslopeI + 
        \bslopeII \right]} {2\int d^2\bt d^2\rt K_1(\Qbar\rt)\left[\bslopeI+\bslopeII\right]}.
\end{equation}
 Being the root mean square average in $\bt$ space, $B$ is sensitive to the
 $\bt$ dependence of the opacities $\Omega_q$ and $\Omega_g$, which depend on
 $S(\bt)$.

The profile functions $S(\bt)$ are connected to the form factors via Fourier
 transforms. We have investigated three different form factors listed below,
 and compared the results with the relevant experimental data on
the $J/\psi$
 forward slope \cite{Bslope}.
 \begin{enumerate}
 \item An exponential form factor, 
 \begin{equation}\label{expform}
 F_{exp}(t)=e^{\frac{1}{4}R^2t},
\end{equation}
 which transforms into a Gaussian in  impact parameter space. 
 \item An electromagnetic form factor 
 \begin{equation}\label{dipform}
F_{dipole}(t)=\frac{1}{(1-\frac{1}{8}R^2t)^2},
\end{equation}
%
its Fourier transform in $\bt$ space is 
\begin{equation}\label{dipS}
S(b_{\perp})=\frac{1}{\pi 
R^2}\frac{\sqrt{8}\bt}{R}K_1(\frac{\sqrt{8}\bt}{R}).
\end{equation}
\item For completeness, we have also investigated the form factor, suggested
 in \cite{DLcharm},
\begin{equation}\label{DLform}
F_{DL}(t)=\frac{4m_p^2-2.79t}{4m_p^2-t}\frac{1}{(1-t/0.71)^2},
\end{equation} 
where $m_p$ is the proton's mass. The Fourier transforms of $F_{DL}$ is 
given
by, 
\begin{equation}\label{DLS} 
S(\bt)=\frac{1}{4\pi}\left[a_1\left(K_0(\sqrt{0.71}\bt)-K_0(2m_p 
\bt)\right)+
a_2bK_1(\sqrt{0.71}\bt)\right],
\end{equation}
where the values of $a_1\approx 0.8$ and $a_2\approx 0.33$ were calculated
from the numerical coefficients in (\ref{DLform}).
\end{enumerate}

The various coefficients are determined from the normalization condition
$\int d^2 \bt S(\bt) \; = \; 1$. 
\begin{figure}
\begin{center}
\epsfig{file=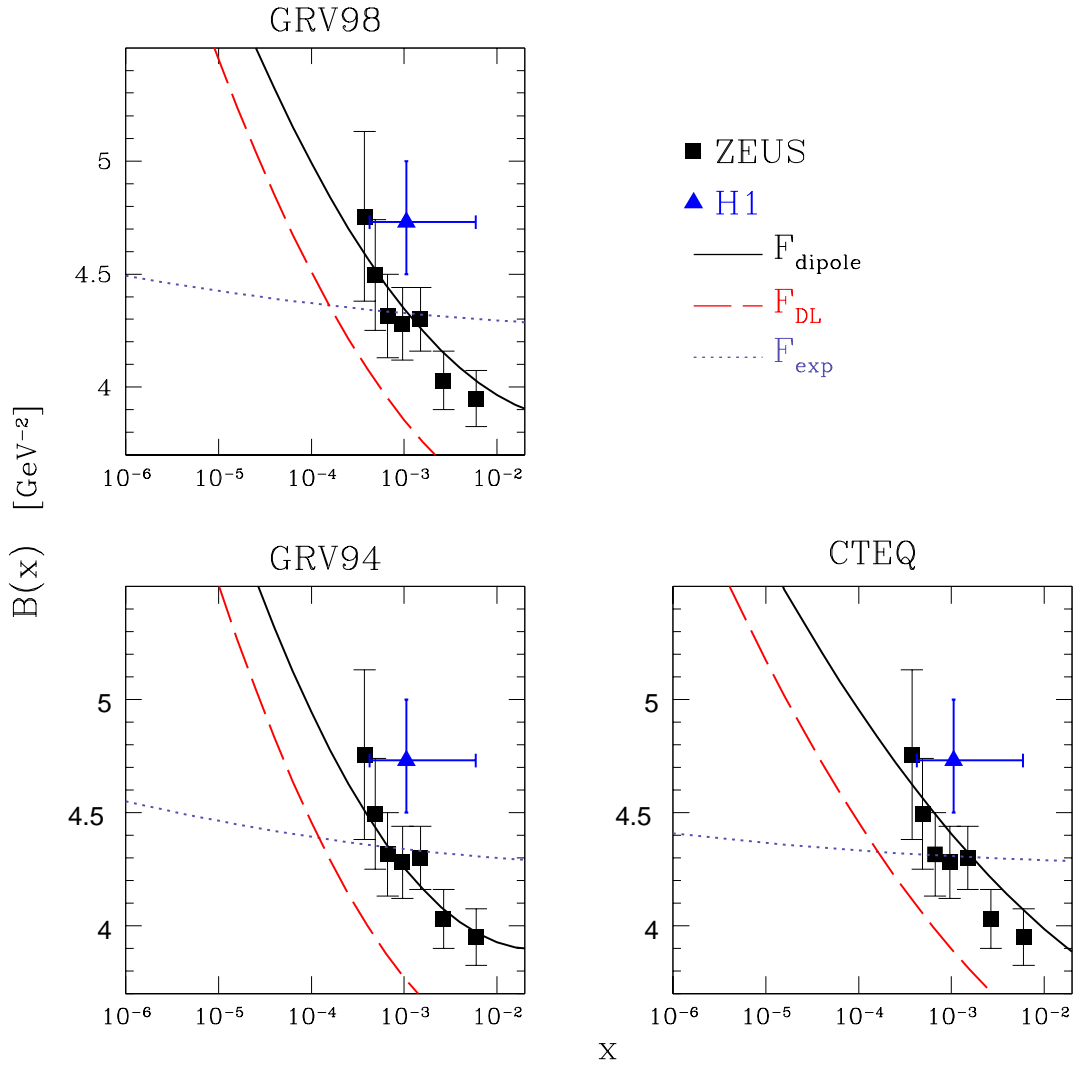,width=15cm}
\end{center}
  \caption[]{\parbox[t]{0.80\textwidth}{\small
The forward slope for differential cross section of $J/\psi$ production, 
data
and predictions, corresponding to different profile functions
$S(\bt)$. Predictions have been calculated for three different PDFs.}}
\label{fig:B}
\end{figure}

In our calculations we have substituted each of the above mentioned profiles
 in (\ref{bslope}), and calculated $B$ for the three different PDF
 parameterizations \cite{GRV94,GRV98,CTEQ5} for the gluon distribution which
 appears in (\ref{omegaq}).  Our results are shown in \fig{fig:B}. While the
 energy dependence obtained using $F_{dipole}$ and $F_{DL}$ is in agreement
 with the ZEUS data, the Gaussian distribution predicts an increase with
 energy which is too mild to reproduce the measured slope.  Regarding
 normalization, the data is well reproduced by (\ref{dipS}) with the choice
 $R^2=10 \; GeV^{-2}$. However, some adjustments are needed regarding the
 numerical parameters of (\ref{DLS}).  The choice of our single parameter, is
 comparable with our earlier estimations ($R^2 = 8.5 \; GeV^{-2}$), where we
 used a Gaussian distribution for $S(\bt)$, and we maintain the quality of
 our results presented in Ref.~\cite{GLMFNpsi}.

\section{\label{sec:low-t} Differential Cross Section at Low ${\mathbf\abst}$}

In this section we calculate $d\sigma/dt$ as a function of $t$ for the region
 of $\abst<Q_0^2$, where $Q_0^2$ is the scale at which the input for the
 evolution is computed.  Our expressions are based on a simplified model of
 two Pomeron exchange, which we find to be satisfactory, given the sizeable
 experimental errors, although our screening correction formalism is capable
 of deriving a full series calculation.  It should be appreciated that what
 follows in this section is a very simplified picture of screening
 corrections to the lowest order and it should not be used for calculating
 other physical quantities, \eg integrated cross sections, on which more
 accurate data exist.

The procedure for calculating the forward differential cross section for
 photo production of a heavy vector meson in the color dipole approximation
 is straightforward \cite{HeavyMeson}.  We follow \cite{GLMFNpsi} and write
 the differential cross section at $t=0$, in the leading logarithmic
 approximation of pQCD, including a contribution from the real part of the
 production amplitude \cite{EDEN} and the skewed (off diagonal) gluon
 distribution,
\begin{equation}\label{pQCD}
\frac{d\sigma(\gamma p\rightarrow V p)}{dt} =
\frac{16\pi^3\Gamma_{ee}}{3\aem M_V^2}\,
\(\frac{2^{2\lambda+3}\,\Gamma(\lambda+\frac{5}{2})}
{\sqrt{\pi}\,\Gamma(\lambda+4)}\)^2\,
\left\{1+\tan^2(\frac{\pi\lambda}{2})\right\}\,\as^2(\Qbar)xG^2(x,\Qbars),
\end{equation}
 where $\lambda=\dlnxg$.  In the derivation of (\ref{pQCD}), a simple static
 non relativistic estimate of the vector meson wave function was used. As
 stated \cite{GLMFNpsi}, we consider the correction due to Fermi motion, which
 changes the numerical value obtained from the above equation, by an overall
 (constant) suppression factor.

In \cite{GLMFNpsi} we took into account further suppression of (\ref{pQCD})
 due to screening. These are facilitated by damping factors \cite{GLMN1,GLMV}
 which are the ratios of the screened to non screened
 observables. Specifically, the DLLA contributions to the damping factors,
 are due to screening in the quark sector (denoted $D_q$) and and in the
 gluon sector (denoted $D_g$). The DLLA screening correction calculation, was
 derived in \cite{GLMV} using a Gaussian distribution for the profile
 $S(\bt)$ [see \eq{expform}].  For completeness, we briefly described the
 basic formulae.

The damping factor due to the screening in the quark sector \ie the
 percolation of the $c \bar c$ through the target, is:
\begin{equation}\label{SCL}
D_{q}^2 =
\(\frac{1}{\kappa_q}E_1(\frac{1}{\kappa_q})e^{\frac{1}{\kappa_q}}\)^2,
\end{equation}
where $\kappa_q$ is given by 
\begin{equation}\label{kappaq}
\kappa_q\,=\,\frac{2 \pi \alpha_S}{3 R^2 {\bar Q}^2}
xG^{DGLAP}(x,{\bar Q}^2).
\end{equation}
Our expression for the damping in the gluon sector, is the square
of the gluon damping defined in \cite{GLMN1,GLMNslope2},
\begin{equation}
\label{xgSC1}
xG^{SC}(x,Q^2)\,=\,D_g(x,Q^2) xG^{DGLAP}(x,Q^2),
\end{equation}
where
\begin{equation}\label{xgSC2}
xG^{SC}(x,Q^2)\,=\,
\frac{2}{\pi^2}\,\int_{x}^{1}\,\frac{dx^{\prime}}{x^{\prime}}
\int_{0}^{Q^2}\,dQ^{\prime\,2}\,
\int db^2\,\(1\,-\,e^{- \kappa_g(x^{\prime},Q^{\prime\,2};b^2)}\),
\end{equation}
and $\kappa_g=\frac{9}{4}\kappa_q$.  The overall damping factor for
(\ref{pQCD}) is $D^2(x,Q^2)=D_q^2(x,Q^2)D_g^2(x,Q^2)$.

\begin{figure}
 \begin{center}
\epsfig{file=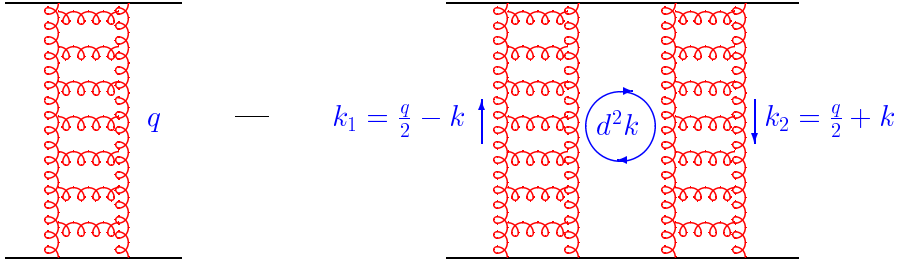,width=14cm,bbllx=41,bblly=563,bburx=636,bbury=707}
\end{center}
  \caption[]{\parbox[t]{0.80\textwidth}{\small
The first two contributions to the amplitude.}}
\label{fig:twopom}
\end{figure}

Since the opacities used in (\ref{SCL})-(\ref{xgSC2}) for calculating
 $D^2(x,Q^2)$ were calculated using a Gaussian profile, we examined the
 sensitivity of $D^2(x,Q^2)$ to the choice of $S(\bt)$, so as to validate our
 normalization for the $t=0$ differential cross section which had been used
 for calculating the integrated cross section.  The result of our
 calculations show that replacing a Gaussian profile with (\ref{dipS}) leads
 to a change of $D^2$ which is less than 4\percent. We thus conclude that as
 far as screening corrections are concerned, we can safely rely on our
 previous estimations at $t=0$.

\eq{eq1} and \eq{eq2} can be viewed as the exchange of many ``hard'' Pomerons
 (gluon ladders). In the region of small $|t|$ we approximate
 this process by the
 exchange of one and two ``hard'' Pomerons (see \fig{fig:twopom}). The
 $t$-dependence of a single Pomeron exchange amplitude is
 proportional to $e^{\half B(W)t}$. The exchange of two ``hard'' Pomerons leads
 to an integration over $d^2k$:

$$
\int\,d^2 k \,\,e^{\half B(W)\left( 
(\frac{\vec{q}}{2} \,-\,\vec{k})^2 \,\,+\,\,
(\frac{\vec{q}}{2} \,+\,\vec{k})^2\,\right)}\,\,\rightarrow\,\,
e^{\frac{1}{4}B(W)t}
$$
where $t = - q^2$.  Hence, in \fig{fig:twopom} the first diagram is
proportional to $e^{\half B(W)t}$ and the second diagram is proportional to
$e^{\frac{1}{4}B(W)t}$.

We therefore approximate the differential cross section as the sum of these
two exchanges \ie,
 \begin{equation}\label{dsdt1}
 \frac{d\sigma(W,t)}{dt}=A^2_{el}\(e^{\half B(W)t}-\eta
 e^{\frac{1}{4}B(W)t}\)^2.
\end{equation}
 We fix our normalization factors $A^2_{el}$ and $\eta$ by the requirement
 that at $t=0$, \eq{pQCD} and \eq{dsdt1}, once screening corrections as well
 as Fermi suppression are taken into account, match. Comparing (\ref{pQCD})
 and (\ref{dsdt1}) including these corrections we find:
\begin{eqnarray}
A^2_{el}(x,\Qbars) & = &
K_f
\frac{16\pi^3\Gamma_{ee}}{3\aem M_V^2}\,
\(\frac{2^{2\lambda+3}\,\Gamma(\lambda+\frac{5}{2})}
{\sqrt{\pi}\,\Gamma(\lambda+4)}\)^2\,
\left\{1+\tan^2(\frac{\pi\lambda}{2})\right\}\,\as^2(\Qbars)xG^2(x,\Qbars) ,
\nonumber\\
\eta(x,\Qbars) &=&1-D(x,\Qbars)\,,
\label{eta}
\end{eqnarray}
where $K_f$ is the Fermi motion normalization factor.

\begin{figure}
\begin{center}
\begin{tabular}{c}
$\uparrow$\\
\mbox{}\\
{${\displaystyle\frac{d\sigma(W,t)}{dt}}${\small $[nb/\gevs]$}}
\end{tabular}
\begin{tabular}{r}
\epsfig{file=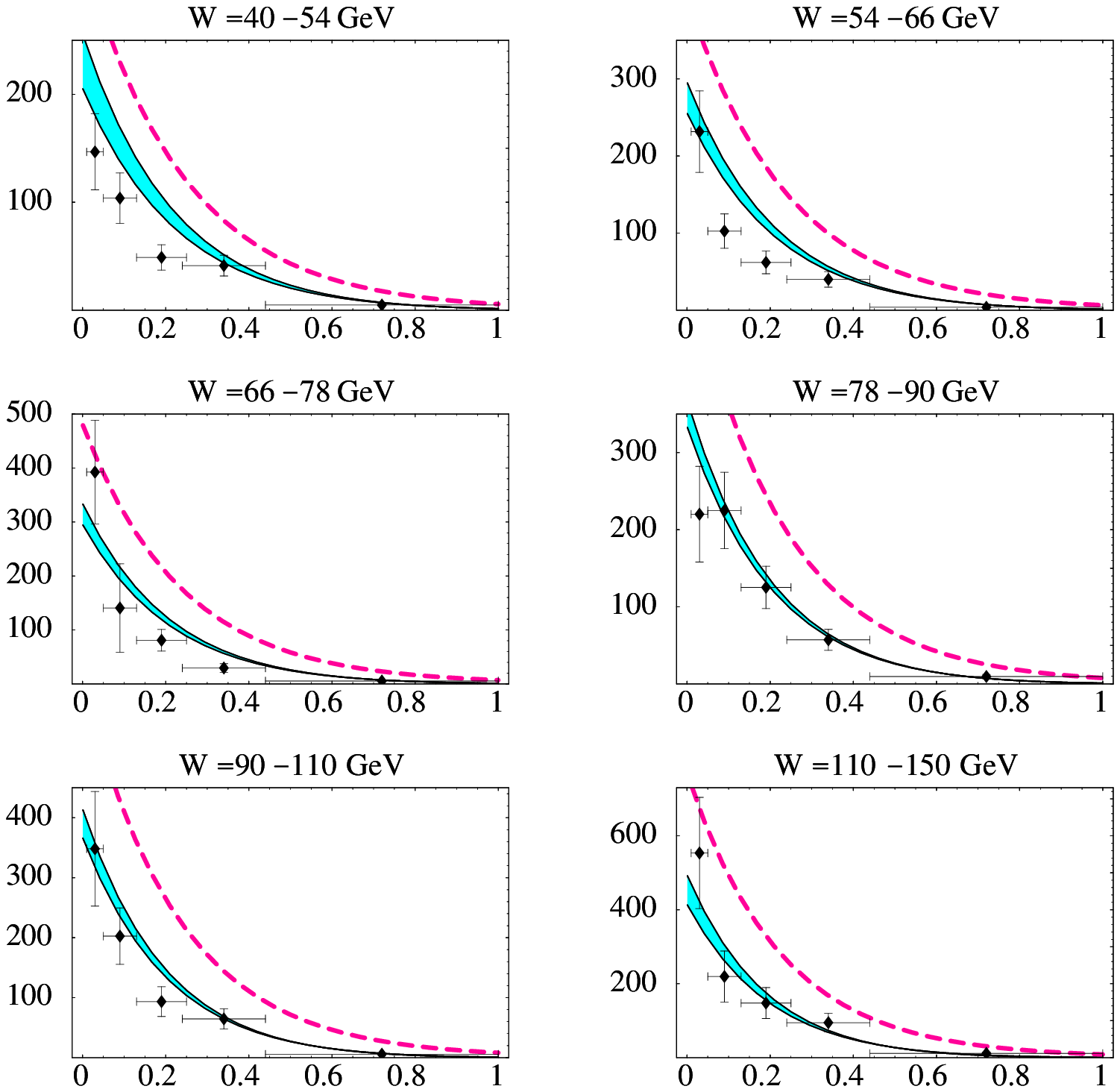,width=12cm,bbllx=100,bblly=180,bburx=540,bbury=620}
 \end{tabular}
 \centerline{\hspace{1cm}$-t\,$ {\small $[\gevs]$ } $\,\,\rightarrow$}
\end{center}
  \caption[]{\parbox[t]{0.80\textwidth}{\small
Elastic differential cross section at small $t$ for different energy ranges.
The shaded areas correspond to each energy range, as given by
Ref.~\cite{H1lowt}, the dashed lines correspond to a single-term exponential
form, as further detailed in the text.}}
\label{fig:lowt}
\end{figure}

We calculate $d\sigma/dt$ using (\ref{dsdt1}) and (\ref{eta}) for low values
 of $t$, and compare the numerical calculations with the experimental data
 for elastic exclusive $J/\psi$ production \cite{H1lowt}. Note that our
 model, as presented in (\ref{dsdt1})-(\ref{eta}), contains no additional
 parameters.  Specifically, the normalization at $t=0$ is taken from the
 integrated cross section data \cite{GLMFNpsi}, and $B$ is taken from section
 \ref{sec:B}.  Our results are shown in \fig{fig:lowt}. The shaded areas in
 \fig{fig:lowt} correspond to the numerical calculations of (\ref{dsdt1}) at
 each energy range given by Ref.~\cite{H1lowt}, and the dashed lines were
 calculated by putting $\eta=0$ in (\ref{dsdt1}).  It can be seen that a
 single-term exponential expression cannot describe the experimental data
 without modification to the overall normalization, which, as stated, is taken
 from rich data of integrated cross section. On the other hand, the
 reproduction of the data is reasonably good, when considering both terms of
 \eq{dsdt1}.

It is interesting to examine numerically the relative role of the first and
 second contributions of (\ref{dsdt1}). Squaring the ratio between the second
 term and the first term we have $\eta^2e^{\frac{1}{2}B(W)|t|}$, which is not
 only an increasing function of $|t|$ but also an increasing function of $W$,
 since both $B$ and $\eta$ increase with the energy.  Hence, the second
 term becomes more and more important as $\abst$ and/or $W$ increase.
 According to our calculations, at $t=0$ the contribution of the second term
 to the differential cross section varies from 4\percent at $W=50\gevs$ to
 10\percent at $W=150\gevs$.  At $\abst\approx Q^2_0$ both the first and the
 second terms of (\ref{dsdt1}) are numerically larger than the value of
 $d\sigma/dt$, where the second term is $20(60)$\percent from the first term
 at low (high) center of mass energy.

As the theoretical uncertainties are difficult to estimate, we used the
 experimental errors to calculate the deviation of our simplified model from
 the data. The corresponding $\chi^2/\mbox{n.d.f}$ for the entire measured
 energy range is 0.7.  As in \cite{GLMFNpsi}, our parameters are strongly
 constrained by the experimental data.

 \section{\label{sec:high-t} Differential Cross Section at Intermediate
   ${\mathbf\abst}$} 

The inelastic contribution to the cross section becomes more and more
 important as the momentum transfer increases. The exponential $t$ dependence
 (\ref{dsdt1}) was written using the elastic forward slope derived in section
 \ref{sec:B}, which is compatible with the experimental data. In this section
 we would like to suggest a simple expression for the inelastic differential
 cross section.

As stated in the introduction, hard processes can be calculated using the
 DGLAP equation, provided the scales of the interaction are such that the
 momenta on the gluon ladder are ordered.  The production of a heavy vector
 meson can be viewed as the exchange of a gluon ladder between a nucleonic
 target and vector meson, which is a $q\bar{q}$ system of small transverse
 size ($\sim 1/M_V^2$).  For intermediate values of $\abst$, the momenta
 are still ordered on the ladder, hence we argue that the DGLAP evolution
 continues to play a role up to $\abst$ of the order of $M_V^2$.
 Specifically, 
 \begin{figure}[t]
 \begin{center}
\epsfig{file=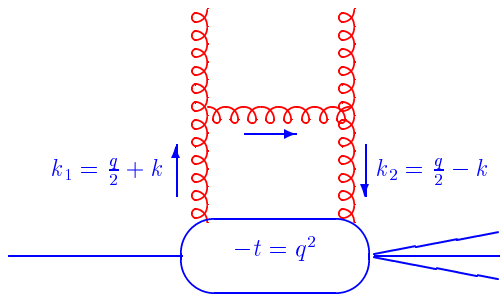,width=7cm,bbllx=90,bblly=480,bburx=340,bbury=638}
\end{center}
  \caption[]{\parbox[t]{0.80\textwidth}{\small
The proton vertex at $-t>Q_0^2$.}}
\label{fig:rang}
\end{figure}
 we can write the amplitude of the lowest rung of the ladder at the proton
 vertex (see \fig{fig:rang}) using the two gluons proporgators $k_1^{-2}$ and
 $k_2^{-2}$.  The corresponding amplitude is proportional to:
\begin{equation}\label{rang}
\frac{(k_1)\cdot(k_2)}{(k_1)^2(k_2)^2},
\end{equation}
 where $k_1=(q/2 + k)$, $k_2=(q/2 - k)$ and the factor $k_1\cdot k_2$
 is implied by gauge invariance. From (\ref{rang}) one can see that the
 dominant contribution to the evolution equation comes from large logarithms
 which appear when $k^2>q^2/4=-t/4$.  For $\abst$ which is not too
 large, \ie in the intermediate region, the DGLAP equation is still valid for
 calculating the production amplitude utilizing a gluon distribution function
 which evolves from $-t/4$ rather than from $Q_0^2$.

 \begin{figure}[t]
\centerline{\hspace{3cm}$W\approx 100\gev$}
\vspace{1.5cm}
\centerline{${\displaystyle\frac{d\sigma(x,\Qbars,t)}{dt}}$}
\vspace{-3cm}
 \begin{tabular}{c}
{\small$[nb/\gevs]$} \\ \vspace{5cm}
\end{tabular} 
\begin{tabular}{l}
\epsfig{file=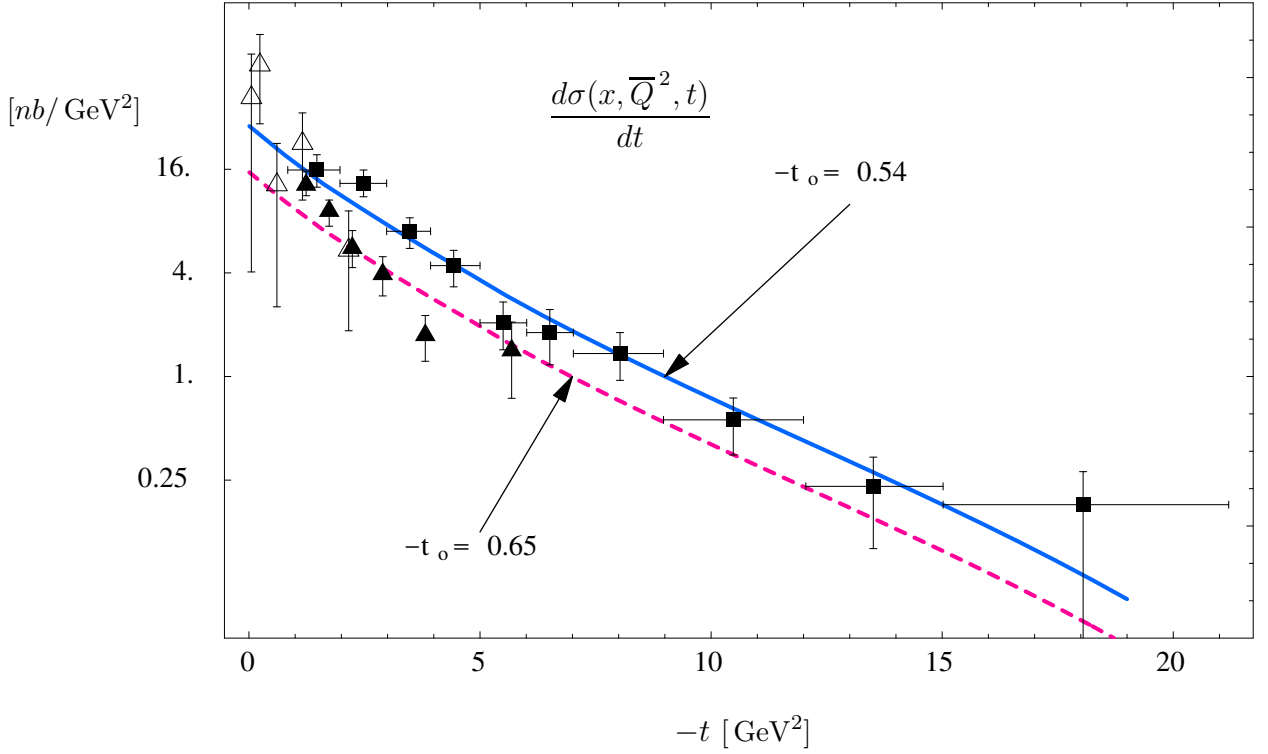,width=12cm,bbllx=163,bblly=245,bburx=540,bbury=570}
\end{tabular}
\centerline{\hspace{3cm}$-t\,\,${\small $[\gevs]$}}
  \caption[]{\parbox[t]{0.85\textwidth}{\small
Inelastic differential cross section in the intermediate region of
$\abst$. Our calculations are presented, for two different values of the free
parameter $t_0$, together with the experimental 
data
of H1 preliminary (squares), ZEUS 1995 (open triangles) and ZEUS 
preliminary
(solid triangles)}} 
\label{fig:hight}
\end{figure}

The physical interpretation of the above discussion is simple.  At small
 $\abst$ ($<1/m_p^2$) the size of the target is roughly the proton size, the
 ``evolution range'' is large and the evolved gluon distribution consists of
 a large number of gluons.  At higher values of momentum transfer the
 effective size of the target is of the order of $1/\abst$. This leads to a
 smaller number of emitted gluons, or in other words to a smaller numerical
 value of $xG$.

From \eq{pQCD} we note that the differential cross section is
 determined by $xG$.  The gluon distribution is related to the
 numerical value of the opacities, thereby to the screening corrections to
 the process.  Thus, the main $t$ dependence of the process comes from the
 scale from which the gluon distribution evolves.

Based on the above discussion, we propose the following expression for the
 differential cross section: 
 \begin{equation}\label{dsdt2}
 \frac{d\sigma(x,\Qbars,t)}{dt}=A^2_{in}D^2(x,\Qbars,t)\, xG^2(x,\Qbars,t)
 \end{equation}
 where,
 \begin{eqnarray}\label{xgt}
 xG(x,\Qbars,t) &=& xG\(x,\frac{\Qbars}{1+\frac{t}{4Q_0^2}}\)\,, \\
 x &=& \frac{Q^2+M_{\psi}^2-t}{W^2}
\end{eqnarray}
 and $D(x,\Qbars,t)$ is calculated by substituting (\ref{xgt}) in
 (\ref{omegaq}) and carrying out the screening corrections procedure
 \cite{GLMFNpsi} in both sectors. 

The justification for choosing the particular dependence of variables that
 the gluon distribution depends on is to incorporate its depenence on $t$.
 Eq.~(\ref{xgt}) is written in a typical dimensional form, \ie the second
 argument of the gluon distribution has dimensions of $\gevs$. This form is
 useful for most of the available parameterizations
 \cite{GRV94,GRV98,CTEQ5}, where each parameterization evolves from a slightly
 different scale $Q_0^2$. Since the DGLAP evolution is manifested
 by logarithms of $Q^2/Q_0^2$, the gluon distribution function has a
 dimensionless form.  Our ansatz may be better understood by considering a
 non-dimensional form of (\ref{xgt}), namely $xG(x,\Qbars/(Q_0^2+\abst/4))$
 were it can be seen that as $\abst$ increases, it dominates the scale of
 evolution whereas at small $\abst$ the gluon distribution function coincide
 with $xG(x,\Qbar/Q_0^2)$.

In (\ref{dsdt2}), the coefficient $A^2_{in}$ is determined by a continuity
 requirement, that (\ref{dsdt1}) and (\ref{dsdt2}) match at $t=t_0$, where
 $t_0$ is our separation parameter. The results of the numerical
 calculations, for two different values of $t_0$, are shown in
 \fig{fig:hight}, together with the H1 and ZEUS data which have been read off
 the plot in \cite{H1larget} and \cite{ZEUSlarget}, respectively.  As can be
 seen from \fig{fig:hight} the large $\abst$ preliminary data set of ZEUS
 \cite{ZEUSlarget} is below the H1 preliminary data \cite{H1larget}.

Our calculations of \eq{dsdt2} depends on the fitted value of our free
 separation parameter $t_0$.  The upper curve shown in \fig{fig:hight}
 corresponds to $|t_0|=0.54 \gevs$, at which $A^2_{in}$ has been determined
 to reproduce the H1 data with $\chi^2/\mbox{n.d.f}=0.65$. However, as the
 preliminary sets of H1 and ZEUS data are consistently different, the task of
 finding a satisfactory description of both sets using the same parameters is
 not trivial. The best $\chi^2$ for the ZEUS data is obtained with
 $|t_0|=0.65 \gevs$ (the lower curve of \fig{fig:hight}) where we calculated
 $\chi^2/\mbox{n.d.f}=1.53$\,. Note that the difference between the two
 fitted values of $t_0$ reflects the presumed difference between the H1 and
 ZEUS sets. This is not necessarily a genuine difference and may just be a
 consequence of different cuts applied at the diffractive proton vertex, by
 the two collaborations.

At first sight, it looks as if (\ref{dsdt2}) may affect the results of our
 previous publication\cite{GLMFNpsi}, {\em inter alia} on account of the
 $t$-dependence of screening corrections at high $\abst$.  However, this
 effect is compensated by the value of the forward slope, which we calculate
 using the dipole electromagnetic form factor, as opposed to an exponential
 form factor used in \cite{GLMFNpsi}.  Indeed, we found that we still
 reproduce all the available experimental data.  The compensation is
 so accurate, that the two curves practically overlap, and we have the same
 $\chi^2$ values as given in \cite{GLMFNpsi}.

\section{\label{sec:conclusions} Summary and Conclusions}

In this paper we have explored the momentum transfer dependence of exclusive
 production of $J/\psi$ vector mesons. By using a Fourier transform of an
 electromagnetic form factor as the profile function in the impact parameter
 space, we calculated the forward differential cross section slope.

We used the expression derived for the slope and calculated the differential
 cross section for the elastic production of $J/\psi$ by using a simple
 exponential expression  approximating one and two Pomeron exchange.

For the inelastic process, we calculated the differential cross section as a
 function of $t$, based on a somewhat naive picture of the interaction. In
 which we suggested replacing the argument of the gluon distribution so as
 to obtain a decreasing function of $\abst$, which coincides with pQCD
 calculations at small $\abst$.

Our conclusions are:
 \begin{enumerate}
 \item The value of the $B$ slope is  sensitive to the choice of the
profile function $S(\bt)$. Our calculations of $B$ reproduce the experimental
data and are not sensitive to the choice of the PDF. 
 \item The fact that the screening correction formalism with a dipole-type
$b_{\perp}$ dependence ($F_{dipole}$) reproduces both the value and the
energy dependence of the $t$-slope, is important for the understanding of the
energy dependence of the ``hard'' Pomeron trajectory. The contributions of
the screening corrections increase with energy and therefore $\alpha'_{eff}$,
the effective slope of the ``hard'' Pomeron trajectory, should also increase
with energy.  This result is particular to a screening corrections
approach and is not obtained in non-screened pQCD calculations.
 \item The screening corrections, which we express through the damping factor
$D$ are not effected by a change in the $b$ dependence of the opacities.
 \item An exponential form of the differential cross section, which is in
agreement with the experimental fits to the forward slope, is suitable for
reasonable description of the elastic, low $\abst$, experimental data. This
form consists of the first two terms of a multi Pomeron exchange, which are
supplemented by screening corrections, as well as corrections due to Fermi
motion of the quarks within the charmonium system. This approximation is also
valid for describing the integrated cross section.
 \item At intermediate values of $\abst$, inelastic data are well reproduced
 by pQCD which is based on DGLAP evolution. The value of $\abst$ at which the
 inelastic processes are comparable with elastic processes is $-t_0\approx
 0.5\gevs$.
 \item At intermediate values of $\abst$ the screening corrections dependence
   on $t$ reflects the decrease of the gluon distribution function.

\end{enumerate}

\section*{Acknowledgments}

We wish to thank E.~Ferreira, who participated in the preliminary stages of
 this investigation, for his contribution and comments.  E.G. wishes to
 acknowledge the hospitality of the Theoretical Nuclear Physics Group at BNL,
 where part of this work was done.  E.L.\ thanks the Theory Group at DESY for
 the stimulating environment provided while U.M.\ is indebted to LAFEX at
 CBPF (Rio de Janeiro) for their support.  This research was supported by the
 BSF grant \# 9800276, by the GIF grant \# I-620-22.14/1999 and by the Israel
 Science Foundation, founded by the Israel Academy of Science and Humanities.

\newcommand{\refbrake}{\\\hspace*{2mm}}

\end{document}